\documentclass[journal]{IEEEtran}
\usepackage{epsfig,setspace,amsmath,epsf,amssymb,bm,theorem,cite,graphicx, epstopdf,algorithm,algpseudocode,float,color,mathtools,tikz,booktabs,bm,physics,arydshln}
\usetikzlibrary{positioning}
\usetikzlibrary{decorations.text}
\usetikzlibrary{decorations.pathmorphing}

\IEEEoverridecommandlockouts
\allowdisplaybreaks

\title{Age of Information in a Single-Source Generate-at-Will Dual-Server Status Update System}
\author{Nail~Akar,~\IEEEmembership{Senior~Member,~IEEE,}
	and~Sennur~Ulukus,~\IEEEmembership{Fellow,~IEEE}
	\thanks{N.~Akar is with the Electrical and Electronics Engineering Department, Bilkent University, Bilkent 06800, Ankara, Turkey. e-mail: {\tt akar@ee.bilkent.edu.tr}. This work is done when N.~Akar is on sabbatical leave as a visiting professor at University of Maryland, MD, USA, which is supported in part by the Scientific and Technological Research Council of T\"{u}rkiye  (T\"{u}bitak) 2219-International Postdoctoral Research Fellowship Program.}
    \thanks{S.~Ulukus is with the Department of Electrical and Computer Engineering, University of Maryland, MD, USA. e-mail: {\tt ulukus@umd.edu}}}
\begin{document}
\maketitle
\begin{abstract}
We study age of information (AoI) in a single-source dual-server status update system for the generate at will (GAW) scenario, consisting of an information source, dual servers, and a monitor. For this system, the method of stochastic hybrid systems (SHS) was used to obtain the mean AoI for the work-conserving ZW (zero wait) policy with out-of-order packet discarding at the monitor. In this paper, we propose a non-work-conserving F/P (freeze/preempt) policy for which the sampling and transmission process is frozen for an Erlang distributed amount of time upon each transmission, and out-of-order packets are preempted immediately at the source, rather than being discarded at the monitor upon reception. We use the absorbing Markov chain (AMC) method to obtain the exact distributions of AoI and also the peak AoI (PAoI) processes, for both ZW and F/P policies. Numerical results are presented for the validation of the proposed analytical model and a comparative evaluation of ZW and F/P policies.
\end{abstract}

\section{Introduction}\label{sec:Intro}
\IEEEPARstart{P}{roviding} timely status updates from physical processes to remote monitors or controllers, over wireless communication networks, has become a key research topic of interest for the deployment of successful internet of things (IoT) applications and services \cite{elmagid_commag19, tripathi_phd23}. For example, in autonomous driving, status information from a number of physical processes including velocity, acceleration, position, trajectory, etc. are to be sent in a timely fashion to a network controller for successfully controlling the vehicle \cite{xu_etal_infocom22}. Examples for other well-known status update systems are environment monitoring applications and wireless body sensor networks \cite{abbas_survey23}.  

Designing network protocols, systems and applications for timely status updates requires the quantification of information freshness by suitable metrics. One well-established source-agnostic freshness metric, i.e., one that does not depend on the source dynamics, is derived from the age of information (AoI) process that was introduced in \cite{kaul_etal_infocom12} for quantifying the timeliness of information freshness in status update systems. The AoI for a given information source is a random process denoted by $\Delta(t)= t-g(t)$ where $g(t)$ refers to the generation time of the freshest status update packet received at the destination.  This process increases with unit slope except that it is subject to abrupt drops upon  a fresher (than the ones previously received) packet reception. On the other hand, the peak AoI (PAoI) process $\Phi_l, l \in \mathbb{Z}^+$ is obtained by taking samples from the AoI process at only pre-reception instances \cite{costa_peak}. We refer the reader to \cite{RoyYates__AgeOfInfo_Survey, kosta_etal_survey,abbas_survey23} and the references therein, for recent surveys on AoI-related research. In most existing work, the particular performance metric is chosen to be the mean AoI (or equivalently the time-averaged AoI) for single-source models, or the weighted mean AoI for their multi-source counterparts, whereas the distribution of AoI has also attracted the attention of researchers, but to a much lesser extent \cite{yates_mgf,akar_gamgam_comlet23}.

One basic mechanism to improve information freshness is to use network (or path) diversity for which information sources use multiple independent network routes for transmitting their status update packets to the remote monitor \cite{kam_etal_TIT16}. We study the exact distributions of both AoI and PAoI processes in a status update system involving a single information source composed of a sensor sampling a corresponding random process, two independent heterogeneous servers representative of two communication paths, and a remote monitor. The goal of the source is to keep the information as fresh as possible at the monitor by effectively using both of the servers for transmission of its status update packets. We assume a generate at will (GAW) scenario in which the source decides when to simultaneously sample the process and transmit the associated information packet. We call the system of interest GAW-2, stemming from dual servers. The case of sampling and transmission decisions belonging to two separate agents, where sampling takes place according to a random process from the perspective of the transmission agent, also known as the random arrival (RA) model \cite{hsu_etal_TMC20,akar_gamgam_comlet23}, is left outside the scope of the paper. 

In this paper, we investigate two particular policies. The first policy is the work-conserving ZW (zero wait) policy \cite{chen_etal_comlet23} where the information source always keeps the servers busy by immediate transmission of a fresh packet whenever the servers become available, and the monitor discards the out-of-order packets at the monitor, upon reception. The second policy is a non-work-conserving F/P (freeze/preempt) policy that we propose in this paper. In F/P policy, sampling and transmission process is frozen (or halted) for an Erlang-$k$, $k=1,2,\ldots,$ distributed amount of time upon  each transmission, i.e., {\em freeze} component of the F/P policy. Moreover, out-of-order packets are preempted immediately at the source, rather than being discarded at the monitor upon reception, which is the {\em preempt} component of the F/P policy. These two components are independent and one may enable only one of the two components, for a given scenario, if desired. 

We study both policies (ZW and F/P) with exponentially distributed service times for the two servers with potentially different service rates. In the absence of a freezing duration, it is possible that information packets with timestamps close to each other, would be transmitted over the two servers, which might be wasteful of network resources in terms of AoI. However, large freezing durations are also wasteful of physical resources, giving rise to a trade-off between information freshness and freeze duration. The reason for the particular Erlang-$k$ freezing is two-fold: First, the Erlang-$k$ distribution is suitable as the freezing model for the analytical method making use of absorbing CTMCs, since this distribution is obtained from time to absorption in a CTMC with $k$ transient states. 
Second, as $k \rightarrow \infty$, the Erlang-$k$ distribution converges to deterministic freezing in distribution. Therefore, the analytical method developed for Erlang-$k$ freezing can be used with sufficiently large $k$ in order to determine how long, i.e., deterministically, the transmission process should be frozen as a function of the system parameters. The method of using Erlang-$k$ distributions to mimic deterministic variables for reasons of analytical tractability is known as Erlangization in the applied probability literature \cite{asmussen.erlangian.2002, ramaswami.2008}. 

The contributions of this paper are as follows. 
\begin{itemize}
\item For ZW, we adopt the absorbing Markov chain (AMC) method of \cite{akar_gamgam_comlet23} to obtain the distribution and moments of AoI (in addition to its mean obtained in \cite{chen_etal_comlet23}) and also that of the PAoI process. The obtained distributions are shown to be in matrix exponential form leading to expressions for higher order moments involving matrix inversion, for the GAW-2 system employing ZW policy.  
\item We propose a non-work conserving F/P (freeze/preempt) policy for the GAW-2 system and use the AMC method to derive the exact distribution of AoI for the GAW-2 system employing F/P policy in matrix exponential form, and we show that preemption by the source improves the AoI performance in comparison to ZW. We obtain further performance improvements with freezing provided the mean freezing time is chosen appropriately. 
\item For both ZW and F/P policies, the exact distribution and moments of PAoI are also obtained in 
closed form by employing the AMC method.
\end{itemize}
The remainder of this paper is organized as follows. Section~\ref{section:Related} summarizes the related work. Section~\ref{sec:prel} presents preliminaries needed to follow the paper. In Section~\ref{sec:SystemModel}, detailed system models for the ZW and F/P policies for the GAW-2 system are given. Section~\ref{sec:analytical} presents the analytical models for the ZW and F/P policies, in two separate sub-sections.  Validation of the analytical models and comparative evaluation of the proposed F/P policy is presented in Section~\ref{sec:Numerical}. Conclusions, open problems and future research directions are given in Section~\ref{sec:Conclusions}. 

\section{Related Work} \label{section:Related}
Early results on AoI were obtained in a single-source single-server setting.  \cite{kaul_etal_infocom12} derives the average AoI for various queuing systems under the FCFS discipline, whereas \cite{kaul_etal_ciss12} studies the LCFS variation of the same problem. The authors of \cite{costa_etal_TIT16} study the M/M/1/2$^{\ast}$ queue for which the packet waiting in the queue is to be replaced by a fresh packet arrival. Reference \cite{soysal_ulukus_IT21} considers the average age of information for a G/G/1/1 system with blocking or preemption. The authors of \cite{akar_etal_tcom20} obtain the exact distribution of AoI in a bufferless status update system with probabilistic preemption, and also single-buffer systems with probabilistic replacement of the buffered packet with a newcoming fresh packet. 

There has also been interest in queuing models for multi-source status update systems. Stochastic hybrid systems (SHS) approach was proposed in \cite{yates2019} for obtaining the mean AoI for a single-buffer server handling status update packets randomly arriving from multiple sources. This method has later been extended to obtain the moment generating function (MGF) and also the higher order moments of AoI, in various settings. In \cite{yates_mgf}, differential equations are derived for the temporal evolution of both the moments and the MGF of the age vector components for a variety of status update systems using the SHS approach. The MGF of AoI has been obtained for a bufferless multi-source non-preemptive or globally preemptive status update system \cite{moltafet_etal_tcom22}. In \cite{akar_gamgam_comlet23}, the authors propose the AMC method to obtain the distribution of AoI for both GAW and single-buffer servers, the latter allowing probabilistic replacement of the packet in the waiting room. 

Another line of research related to AoI is the presence of path diversity in a status update system for which there are multiple servers handling the transmission of status update packets from a single source  or from multiple sources. The authors  of \cite{kam_etal_TIT16} study the FCFS M/M/2 dual-server RA model and propose a method to obtain the mean AoI as well as its approximation and lower/upper bounds whereas \cite{yates_isit18} considers a system with multiple paths where each path is modeled as a preemptive LCFS M/M/1/1 queue and derives the mean AoI using the SHS approach. 
Reference \cite{bedewy_etal_isit16} studies a multi-server RA system for which stale packets in the queue are not dropped, and they show that the preemptive last generated first serve (LGFS) policy simultaneously optimizes performance in terms of data freshness, throughput,
and delay when service times are iid and exponentially distributed.
The authors of \cite{javani_etal_globecom19} derive the mean AoI for the LCFS RA model using SHS for a single source and two heterogeneous servers, and also general number of sources with either two or three homogeneous servers. 

The focus of this paper is a GAW status update system with two heterogeneous servers, i.e., GAW-2, which is different than the RA models studied above. The most relevant work to our study is reference \cite{chen_etal_comlet23} which uses SHS to obtain the mean AoI for the GAW-2 system employing the work-conserving ZW policy, i.e., when a transmission is complete on a given server, the successive transmission gets to start immediately, and with out-of-order packets discarded at the monitor, when the service times are exponentially distributed. The authors of \cite{chen_etal_twc23} additionally  study the case when one of the service times is deterministic and they derive closed-form expressions for the average AoI and average peak AoI.

\section{Preliminaries}\label{sec:prel}
In the following, we describe an absorbing Markov chain (AMC), its absorption probabilities, and distribution of time until absorption, based on \cite{kemeny1960finite} and \cite{neuts81}, that are needed to follow the proposed analytical method. 
We use bold letters for vectors and matrices. Consider a continuous-time Markov chain (CTMC) $X(t), t\geq 0,$ with $N$ transient and $M$ absorbing states. Let the generator of this CTMC $X(t)$ be written as,
\begin{align}
\bm{Q}&=
 \left[ \begin{array}{c:c}
   \bm S & \bm V \\
   \hdashline
   \bm{0} & \bm{0} \\
\end{array}\right], \label{eq:Smatrix}
\end{align}
where the ${N \times N}$ matrix $\bm S$ and ${N \times M}$ matrix $\bm V$ are composed of the transition rates among the transient states, and from the transient states to the absorbing states, respectively. 
When all the absorbing states are merged into one single absorbing state, the distribution of time until absorption to any of these absorbing states, denoted by $\Theta$, is known to have a phase-type (PH-type) distribution with probability density function (pdf) in the form,
\begin{align}
    f_{\Theta}(x)=\bm{\sigma} \mathrm{e}^{\bm{S}x}\bm{\nu},
    \label{eq:pdf}
\end{align}
where $\bm{\sigma} = \{ \sigma_j \}$ is a ${1 \times N}$ row vector, $\sigma_j$ denotes the initial probability of being in transient state $j$, $\bm{\nu}= \bm{V} \bm{1} = - \bm{S} \bm{1}$ and $\bm{1}$ denotes a column vector of ones of appropriate size.
In this case, we say $\Theta \sim \text{PH}(\bm{\sigma},\bm{S})$ of order $N$, i.e., $\Theta$ is characterized with 
a vector $\bm{\sigma}$, and a matrix $\bm{S}$, with $N$ being the size of the characterizing matrices. The states are also called {\em phases} of a PH-type distribution.
The cumulative distribution function (cdf) of $\Theta$ can be written as,
\begin{align}
    F_{\Theta}(x)=\bm{\sigma} (\mathrm{e}^{\bm{S}x} - \bm{I}) \bm{S}^{-1} \bm{\nu}.
    \label{eq:cdf}
\end{align}
Following the expression \eqref{eq:pdf}, the $i$th non-central moment of $\Theta$ is written as, 
\begin{align}
   \mathbb{E}[\Theta^i] & = \int_{x=0}^{\infty} x^k f_{\Theta}(x) \dd{x} =
   (-1)^{i+1} i! \bm{\sigma} \bm{S}^{-(i+1)} \bm{\nu}.
\end{align} 
In particular,
\begin{align}
   \mathbb{E}[\Theta] & = \bm{\sigma} \bm{S}^{-2} \bm{\nu} = 
    -\bm{\sigma} \bm{S}^{-1} \bm{1}. \label{firstmoment}
\end{align} 
 The probability of absorption in state $m$, $m=1,\ldots,M$, is given by,
\begin{align}
    p_{m} & =-\bm{\sigma} \bm{S}^{-1} \bm{V}_{m}, \label{eq:prob}
\end{align}
where $\bm{V}_{m}$ is the $m$th column of $\bm{V}$. An exponentially distributed random variable $X$ with mean $\lambda^{-1}$ is PH-type of order 1, i.e., $X \sim \text{PH}(1,\lambda)$.
An Erlang-$k$ distributed random variable $X_k \sim \text{Erl}(\lambda,k)$  with mean $\lambda^{-1}$ is the sum of $k$ independent exponentially distributed random variables each with mean $\frac{1}{k\lambda}$ with $\text{Var}(X_k) = \frac{1}{k\lambda^2}$, and hence as $k \rightarrow \infty$, $X_k$ converges to a deterministic variable in distribution. 
For the Erlang-$k$ distribution, we are in phase 1 initially, and when the corresponding AMC $X(t)$ is in phase $\ell < k$ (resp. phase $k$), then we transition to phase $\ell+1$ (resp. absorbing state) both with intensity $k \lambda$, and the time until absorption is said to possess an Erlang-$k$ distribution with mean $\lambda^{-1}$.
Thus, $X_k \sim \text{PH}(\bm{\sigma}_k,\bm{S}_k)$ with order $k$ where 
\begin{align*}
\bm{\sigma}_k & = \begin{bmatrix} 1 & 0 & \cdots & 0  \end{bmatrix},  
\end{align*} and
\begin{align}
\bm{S}_k & = 
\begin{bmatrix} 
-k \lambda  & k \lambda & & & \\
& -k \lambda  & k \lambda & & \\
& & \ddots & \ddots & \\
& & & -k \lambda  & k \lambda \\
& & & & -k \lambda 
\end{bmatrix}.
\end{align}

\section{System Model} \label{sec:SystemModel}
\begin{figure}[tb]
    \centering
    \begin{tikzpicture}[scale=0.3]
    \pgfplothandlercurveto
    \pgfplotstreamstart
    \pgfplotstreampoint{\pgfpoint{0.5cm}{4.cm}}
    \pgfplotstreampoint{\pgfpoint{1cm}{5.75cm}}
    \pgfplotstreampoint{\pgfpoint{1.5cm}{7.2cm}}
    \pgfplotstreampoint{\pgfpoint{2cm}{6.7cm}}
    \pgfplotstreampoint{\pgfpoint{2.5cm}{5.7cm}}
    \pgfplotstreampoint{\pgfpoint{3cm}{6.9cm}}
    \pgfplotstreampoint{\pgfpoint{3.5cm}{5.2cm}}
    \pgfplotstreampoint{\pgfpoint{4cm}{5.7cm}}
    \pgfplotstreampoint{\pgfpoint{4.5cm}{5.2cm}}
    \pgfplotstreamend
    \pgfusepath{stroke}
    \filldraw[fill=black!1, color=lightgray] (13,1) rectangle(15,3);
    \draw[thick, black] (16,2) circle (1) ;
    \draw (16,2) node[anchor=center] {$\mu_2$} ;
    \draw[lightgray] (11,1) -- (13,1);
    \draw[lightgray] (11,3) -- (13,3);
    \filldraw[color=lightgray] (13,8) rectangle(15,10);
    \draw[thick, black] (16,9) circle (1) ;
    \draw (16,9) node[anchor=center] {$\mu_1$} ;
    \draw[lightgray] (11,8) -- (13,8);
    \draw[lightgray] (11,10) -- (13,10);
    \draw[thick,black,->] (18,2.5) -- (21,4.5) ;
    \draw[thick,black,->] (18,8) -- (21,6) ;
    \filldraw (24,5.5) node[anchor=center] {\small{monitor}};
    \draw[rounded corners,darkgray,thick] (22,4.5) rectangle (26,6.5) {};
    \draw[rounded corners,thick,darkgray,dashed] (0,3.5) rectangle (5,7.5) {};
    \draw[thick] (5,5.5) -- (6,5.5);
    \draw[thick,->] (6,5.5) -- (7.5,7);
    \draw[thick, ->,dotted] (6.5,6.5) arc (80:10:1.2);
    \draw[thick] (7.5,5.5) -- (8.5,5.5);
    \draw[thick,black,->] (9,5.5) -- (10.75,7.5);
    \draw[thick,black,->,dotted] (9,5.5) -- (10.75,3.5);
    \filldraw (2.4,8.25) node[anchor=center] {\small{random process}};
    \filldraw (6.7,4.75) node[anchor=center] {\small{sensor}};
    \filldraw (13.5,0.25) node[anchor=center] {\small{server 2}};
    \filldraw (13.5,7.25) node[anchor=center] {\small{server 1}};
    \end{tikzpicture}
	\caption{A status update system with one source, two heterogeneous servers, and a monitor. In ZW, sampling/transmission takes place upon a service completion, whereas it is governed by the freeze policy in F/P.}
	\label{fig:SystemModel}
\end{figure}
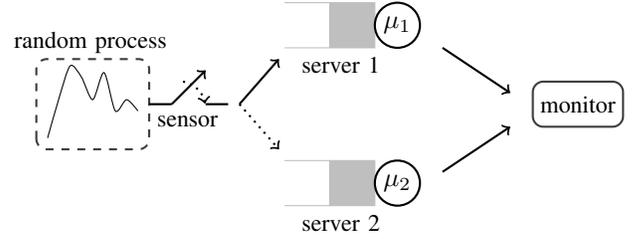

We consider a single information source that samples a random process and generates an information packet containing the sample value, at will. There are two servers handling the transmission of the source packets to the monitor, termed as the GAW-2 model. The service time of server 1 (resp.~2) is exponentially distributed with parameter $\mu_1$ (resp.~$\mu_2$) and without loss of generality we assume $\mu_1 \geq \mu_2$. When a transmission is over, the source is immediately acknowledged. Packet errors during transmission are not assumed in this work. The GAW-2 system is illustrated in Fig.~\ref{fig:SystemModel}.

In the zero wait (ZW) policy studied in \cite{chen_etal_comlet23}, the source immediately feeds any one of the servers with a fresh status update packet immediately when it becomes available for transmission. Therefore, ZW is work-conserving, i.e., the server never idles. Since the two servers are independent, it is possible that packets can reach the destination out-of-order. In ZW, in order for a received information packet to be accepted at the monitor, it needs to have been sampled at a later time than the most recently accepted packet, since otherwise it would not reduce the AoI. Otherwise, this out-of-order packet is discarded. 

We propose to have the following modifications on ZW with the F/P (freeze/preempt) policy introduced in this paper:\begin{itemize}
    \item Whenever a transmission starts on one of the servers, the sampling and transmission process is frozen for an Erlang-$k$ duration $X_k$  which reduces to exponential freezing for $k=1$ and deterministic freezing for $k \rightarrow \infty$. The motivation behind freezing stems from the observation that, if two packets with close-by timestamps are to be transmitted over the two servers, one of these packets, even in case it is not discarded, will have a low contribution to the AoI at the monitor. As a consequence of freezing, F/P is non-work conserving, i.e., servers may occasionally be idle in F/P.
    \item If both servers turn out to be available when a freeze period ends for F/P, the newly generated status update packet is sent over the faster server 1. This situation does not arise for ZW. 
    \item When a packet is acknowledged with a time stamp later than the one in service, then the packet in service becomes obsolete. In the F/P policy, an obsolete packet is preempted by the source while in service. This is in contrast to the ZW scheme of \cite{chen_etal_comlet23}, where an obsolete packet is discarded at the monitor after its transmission is complete. 
    
\end{itemize} 

\begin{figure}[tb]
    \centering
    \begin{tikzpicture}[scale=0.32]	
    \draw[<->,black] (9,13) -- (16,13);
    \filldraw (12.5,13) circle (0.01) node[anchor=south, thick] {cycle-$l$};
    \draw[thick,->] (0,0) -- (23,0) node[anchor=north] {$t$};
    \draw[thick,->] (0,0) -- (0,14) node[anchor=south] {$\Delta(t)$};
    \draw[ultra thick,red] (4.5,4.5) -- (9,9);
    \draw[dashed,very thin] (4.5,4.5) -- (9,9);
    \filldraw[red] (4,4) circle (3pt);
    \filldraw[red] (3.5,3.5) circle (3pt) ;
    \filldraw[red] (3,3) circle (3pt); 
    \draw (0,5) node[anchor=east] {$u_l$};
    \draw[dotted,gray] (0,9) -- (23,9);
    \draw[dotted,gray] (0,5) -- (23,5);
    \draw[dotted,gray] (0,2) -- (23,2);
    \draw[dotted,gray] (9,12.5) -- (9,0)  node[anchor=north, thick, black] {$d_{l}$};
    \draw[ultra thick,red] (9,9) -- (9,5.2);
    \draw[ultra thick,red] (9.1,5.1) -- (16,12);
    \draw[dashed,black,very thin] (16,12) -- (4,0)  node[anchor=north, thick, black] {$t_{l}$};
    \node at (3.75,0.50) (nodeA) {};
    \node at (8.75,5.5) (nodeB) {};
    \node at (15.75,12.5) (nodeC) {};
    \draw [decoration={text along path,
    text={|\scriptsize|states 1-4 (ZW)},text align={center}},decorate]  (nodeA) -- (nodeB);
    \draw [decoration={text along path,
    text={|\scriptsize|states 5-7 (ZW)},text align={center}},decorate]  (nodeB) -- (nodeC);
    \node at (5,-0.5) (nodeA) {};
    \node at (10,4.5) (nodeB) {};
    \node at (17,11.5) (nodeC) {};
    \draw [decoration={text along path,
    text={|\scriptsize|states 1-10 (F/P)},text align={center}},decorate]  (nodeA) -- (nodeB);
    \draw [decoration={text along path,
    text={|\scriptsize|states 11-14 (F/P)},text align={center}},decorate]  (nodeB) -- (nodeC);
    \draw (0,12) node[anchor=east] {$\Phi_l$};
    \draw (0,2) node[anchor=east] {$u_{l+1}$};
    \draw[dotted,gray] (16,12.5) -- (16,0)  node[anchor=north, thick, black] {$\quad d_{l+1}$};
    \draw[dotted,gray] (0,12) -- (23,12);
    \draw[ultra thick,red] (16,12) -- (16,2.2);
    \draw[ultra thick,red] (16.1,2.1) -- (20,6);
    \draw[dashed,black,very thin] (20,6) -- (14,0)  node[anchor=north, thick, black] {${t_{l+1}} \quad$};
    \filldraw[red] (20.5,6.5) circle (3pt);
    \filldraw[red] (21,7) circle (3pt) ;
    \filldraw[red] (21.5,7.5) circle (3pt); 
    \draw[gray] (9,5) circle (6pt);
    \draw[gray] (16,2) circle (6pt);
    \end{tikzpicture}
    \caption{Sample path of the AoI process $\Delta(t)$. Only generation and reception instances of successful packets are shown.}
    \label{fig:samplepath}
\end{figure}
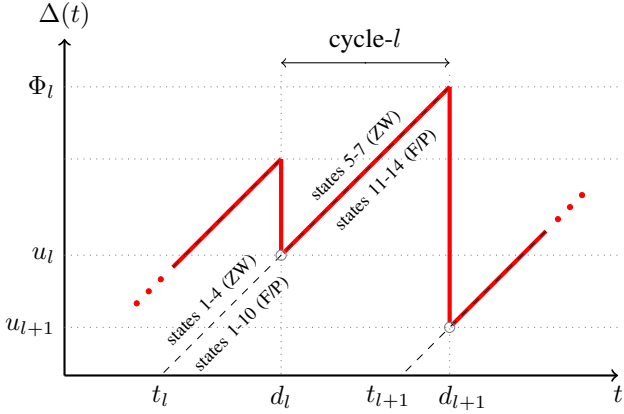

We now describe the AoI process for the GAW-2 system that applies to both ZW and F/P. Let $t_l$ and $d_l$ denote the instances at which the $l$th successful packet is generated by the source and received by the monitor, respectively. Unsuccessful packets are those that are generated but not received by the monitor due to discarding at the monitor (resp.~preemption at the source) for ZW (resp.~F/P). Successful packets are the ones that are accepted by the monitor (resp.~received at the monitor) for ZW (resp.~F/P). Let $u_l$ denote the system time of the $l$th successful packet, i.e., $u_l=d_l - t_l$. Fig.~\ref{fig:samplepath} presents a sample path of the AoI process $\Delta(t)$ (thick red solid curve). During cycle-$l$,  $\Delta(t)$ increases with unit slope from the value $u_l$ at time $d_{l}$ until the value $\Phi_l$ at time $d_{l+1}$ when it drops down to $u_{l+1}$. The random process $\Phi_l, l \geq 1$ is called the peak AoI (PAoI) process. $\Delta$ (resp.~$\Phi$) denotes the steady-state random variable for the random process $\Delta(t)$ (resp.~$\Phi_l$) with cdf $F_{\Delta}$, i.e., $F_{\Delta}(x) = \lim_{t \rightarrow \infty} \mathbb{P} ( \Delta(t) \leq x), \ x \geq 0$, (resp. $F_{\Phi}$, i.e., $F_{\Phi}(x) = \lim_{l \rightarrow \infty} \mathbb{P} ( \Phi_l \leq x), \ x \geq 0$)
and $f_{\Delta}$ (resp.~$f_{\Phi})$ denotes its pdf.

\section{Analytical Models for GAW-2}\label{sec:analytical}

\subsection{Zero Wait (ZW) Policy}

The AMC-based method we propose for GAW-2 is based on \cite{akar_gamgam_comlet23} and is composed of the following three steps. We observe from Fig.~\ref{fig:samplepath} that, a single AoI cycle (e.g., cycle-$l$) starts with the reception of a successful packet (e.g., at time $d_l$) and continues until the reception of the next successful packet (e.g., at $d_{l+1})$. In the first step of the AMC method, we construct an AMC $Y(t)$ with two absorbing states, which starts operation at time $t=0$ (corresponding to $t=t_l$ in Fig.~\ref{fig:samplepath}) with the generation of an arbitrary packet, say $P_{\ast}$. The transient and absorbing states (resp.~transition rates) of this AMC are given in Table~\ref{tab:AMCofZW} (resp.~Table~\ref{tab:TRofZW}). Note that, when this particular packet $P_{\ast}$ is successful (this happens when $t=d_l$ in Fig.~\ref{fig:samplepath}), then the AMC continues evolving until the reception of the next successful packet upon which we reach the successful absorbing state 8 (corresponding to $t=d_{l+1}$ in Fig.~\ref{fig:samplepath}). If this packet is discarded or preempted, the AMC is absorbed into the unsuccessful absorbing state 9.
In these tables, $S_i$ stands for server $i$, $P_i$ stands for the packet transmitted on $S_i$, and $T_i$ stands for the time stamp, i.e., packet generation instance, of the packet in transmission on $S_i$. In fact, we do not need the actual $T_i$ values, but instead we need to keep track of whose time stamp is earlier to identify obsolete packets. A packet in transmission becomes obsolete, when another packet is received with a later time-stamp.  Otherwise, the packet is up to date. Note that $S_i,i=1,2$ is always busy transmitting a packet for the ZW policy due to its work-conserving nature. 

\begin{table}[tb]
\caption{Transient and absorbing states of the AMC process $Y(t)$ for ZW policy.}
\centering
\begin{tabular}{|c|c|} 
 \hline
 State & Description \\
 \hline \hline
 $1$ & $P_{\ast}$ on $S_1$, $T_2 \leq T_1$ \\
 \hline
 $2$ & $P_{\ast}$ on $S_1$, $T_2 > T_1$  \\
 \hline
 $3$ & $P_{\ast}$ on $S_2$, $T_1 \leq T_2$ \\
 \hline
 $4$ & $P_{\ast}$ on $S_2$, $T_1 > T_2$ \\
 \hline
 $5$ &  $P_1,P_2$ up to date \\
 \hline
 $6$ & $P_1$ up to date, $P_2$ obsolete \\
 \hline
 $7$ &  $P_1$ obsolete, $P_2$ up to date \\
 \hline
 $8$ & Successful absorbing state \\
 \hline
 $9$ & Unsuccessful absorbing state  \\
 \hline
\end{tabular}
\label{tab:AMCofZW}
\end{table}

\begin{table}[tb]
\caption{Transition rates for the AMC process $Y(t)$ for ZW.}
\centering
\begin{tabular}{|c|c|c|c|c|c|} 
 \hline
 \multicolumn{3}{|c|}{Transition Rates} & \multicolumn{3}{|c|}{Transition Rates} \\
 \hline \hline
 From & To  & Value & From & To & Value\\ 
 \hline \hline
 $1$ & $6$ & $\mu_1$& $2$ & $5$ & $\mu_1$  \\ \cline{2-3} \cline{5-6}
 & $2$ & $\mu_2$ & & $9$ &  $\mu_2$ \\ \hline 
 $3$ & $4$ & $\mu_1$& 4 & $9$ & $\mu_1 $ \\ \cline{2-3} \cline{5-6}
 & $7$ & $\mu_2$ & & $5$ &  $\mu_2$ \\ \hline 
 $5$ & $8$ & $\mu_1 + \mu_2$ & $6$ & $8$ & $\mu_1 $ \\  \cline{5-6}
 & &  & & $5$ &  $\mu_2$ \\ \hline 
 $7$ & $5$ & $\mu_1$&  \multicolumn{3}{c|}{}  \\ \cline{2-3} 
 & $8$ & $\mu_2$ &  \multicolumn{3}{c|}{}  \\ \hline 
\end{tabular}
\label{tab:TRofZW}
\end{table}

We now explain the transition rates:
\begin{itemize}
    \item When in state 1, either $P_{\ast}$ completes with intensity $\mu_1$, in which case $Y(t)$ will transition to state 6 since at this point $P_2$ becomes obsolete, or $P_2$ completes with intensity $\mu_2$  upon which a transition to state 2 takes place with a fresh packet placed at server 2.
    \item In state 2, either $P_{\ast}$ completes with intensity $\mu_1$, in which case $Y(t)$ will transition to state 5, or $P_2$ completes with intensity $\mu_2$ upon which $P_{\ast}$ becomes obsolete and absorption into state 9 occurs. 
    \item The behaviors in states 3 and 4 are similar to states 1 and 2, respectively, except that $P_{\ast}$ resides in $S_2$ instead of $S_1$.
    \item In states 5, 6, and 7, $P_{\ast}$ has been successfully received and we need one more up to date packet to complete. In state 5, neither of the packets in service is obsolete, and upon a service completion, either on $S_1$ or $S_2$, the AMC is absorbed into state 8.
    \item When in state 6 (resp.~state 7), $P_2$ (resp.~$P_1$) is obsolete and will be discarded with intensity $\mu_2$ (resp.~$\mu_1$), or with intensity $\mu_1$ (resp.~$\mu_2$), the up to date packet following $P_{\ast}$ will be received at the monitor leading to absorption into state 8. 
\end{itemize}

Consequently, in this step, we obtain the  $9 \times 9$ infinitesimal generator $\bm Q$ of the AMC $Y(t)$ as follows, 
\begin{align} 
{\bm Q} & = 
\left[
\begin{array}{ccccccc:cc}
\ast  &  \mu_2   &     0     &    0    &     0 &   \mu_1&        0   &      0  &       0 \\ 
         0  &  \ast   &     0     &    0  &  \mu_1   &    0     &    0    &     0   &  \mu_2 \\ 
         0   &      0 &   \ast  &  \mu_1    &     0   &      0  &  \mu_2   &      0   &      0 \\ 
         0   &      0  &       0 &   \ast &   \mu_2   &     0  &       0  &       0 &   \mu_1\\ 
         0   &      0   &      0  &       0  &  \ast &    0   &      0  & \mu_1 + \mu_2    &      0 \\ 
         0    &     0   &      0  &       0  &  \mu_2 &   \ast  &     0  &    \mu_1   &   0 \\ 
         0    &     0   &      0   &      0   &  \mu_1   &    0  & \ast  &   \mu_2   &    0 \\ \hdashline
0 & 0 & 0 & 0 & 0 & 0 & 0 & 0 & 0    \\ 
0 & 0 & 0 & 0 & 0 & 0 & 0 & 0 & 0  \\
\end{array}
\right] , \label{big1} 
\end{align}
where the
$7 \times 7$ north-west block is denoted by $\bm{S}$, the $7 \times 2$ north-east block is denoted by $\bm{V}$, $\bm{V}_i$, $i=1,2$ denoting the $i$th column of $\bm{V}$, and the $\ast$ symbol should be chosen to make the row sums zero. Note that $\bm{Q}$ is in the same form as \eqref{eq:Smatrix}.

In the second step of the proposed method, we obtain the initial probability vector of the AMC $Y(t)$ which is not difficult to find for the ZW policy since a new packet generation occurs according to a Poisson process with rate $\mu_1 + \mu_2$, i.e., when a server becomes available, and the new packet will be placed at $S_1$ (resp.~$S_2$) with probability $\frac{\mu_1}{\mu_1 + \mu_2}$ (resp. 
$\frac{\mu_2}{\mu_1 + \mu_2}$). Therefore, the initial probability vector, $\bm{\sigma}$, of the AMC $Y(t)$ is given by the $1 \times 7$ row vector $\bm{\sigma}$,
\begin{align}
\bm{\sigma} & = \begin{bmatrix}
\frac{\mu_1}{\mu_1 + \mu_2} & 0 & \frac{\mu_2}{\mu_1 + \mu_2} & 0 & 0 & 0 & 0 \end{bmatrix}.
\end{align}

In the third and final step, we first obtain the pdf of the PAoI, and subsequently that of the AoI process. We observe that the dashed black curve in Fig.~\ref{fig:samplepath} starting at $t_l$ until $d_{l+1}$ amounts to the time spent before absorption for the AMC $Y(t)$ in successful absorption cycles. The portion of this curve from $t_l$ until $d_{l}$ is spent in states 1 to 4 which does not overlap with the AoI curve. On the other hand, the remaining portion of this curve from $d_l$ until $d_{l+1}$ spent in states 5 to 7 is the same as the AoI curve in cycle-$l$.
Revisiting Fig.~\ref{fig:samplepath}, 
the distribution of the time to absorption, denoted by $\Gamma$ of the AMC $Y(t)$, conditioned on successful absorption, overlaps with that of the steady-state PAoI, $\Phi$. Also note that the distribution of $\Gamma$ conditioned on unsuccessful absorption, is not needed at all, for AoI or PAoI distributions. 
Therefore,
\begin{align}
F_{\Phi}(x) & = \mathbb{P} ( \Phi \leq x ), \\ 
& = \mathbb{P} ( Y(x) =8  | Y(\infty) = 8 ), \label{eq:peak1} \\
& = \frac{\mathbb{P} ( Y(x) =8) }{\mathbb{P} ( Y(\infty) = 8)}, \label{eq:peak2}  \\
& = \frac{\bm{\sigma} (\mathrm{e}^{\bm{S}x} - \bm{I}) \bm{S}^{-1} \bm{V}_1}{-\bm{\sigma} \bm{S}^{-1} \bm{V}_1}.
\label{eq:peak3}
\end{align}
The pdf of the peak AoI, $\Phi$, is obtained by differentiating the above expression with respect to $x$, 
\begin{align}
f_{\Phi}(x)  & = \frac{\bm{\sigma} \mathrm{e}^{\bm{S}x} \bm{V}_1}{-\bm{\sigma} \bm{S}^{-1} \bm{V}_1},  \ x \geq 0, \label{density_paoi}
\end{align}
with its mean value given by,
\begin{align}
\mathbb{E} [\Phi] & = \frac{\bm{\sigma} {\bm{S}^{-2}} \bm{V}_1}{-\bm{\sigma} \bm{S}^{-1} \bm{V}_1}. \label{mean_paoi}
\end{align}

Let us now turn our attention to the steady-state AoI, $\Delta$. For this purpose, we visit Fig.~\ref{fig:samplepath} to observe that the probability $\mathbb{P} ( x < \Delta \leq  x + \dd{x} )$ is proportional 
with $\dd{x}$ times the value $x$ is exceeded for the AMC process $Y(t)$, i.e., $\mathbb{P} ( Y(x) \in \mathcal{A})$
with $\mathcal{A}$ being the set of three transient states that overlap with the AoI curve, i.e., 
$\mathcal{A}= \{ 5,6,7 \}$  (see Fig.~\ref{fig:samplepath})
conditioned on absorption into the successful absorbing state 8. Actually, $\mathbb{P} ( x < \Delta \leq  x + \dd{x} )$ is $\mathbb{P} ( Y(x) \in \mathcal{A}| Y(\infty=8)) \dd{x}$ divided by the mean AoI cycle length.
Mathematically, there is a proportionality constant $\kappa$ such that 
\begin{align}
f_{\Delta}(x)  & = \kappa \  \mathbb{P} ( Y(x) \in \mathcal{A}| Y(\infty=8)), \\
               & = \kappa \ \frac{\mathbb{P} ( Y(x)\in \mathcal{A} ) }{\mathbb{P} ( Y(\infty) = 8)}. \label{eq:aoi2}
\end{align}
Noting that the above expression needs to integrate to one, we have 
\begin{align}
f_{\Delta}(x) & = \frac{\bm{\sigma} \mathrm{e}^{\bm{S}x} \bm{\theta}}{-\bm{\sigma} {\bm{S}}^{-1} \bm{\theta}}, \ x \geq 0,\label{density_aoi}
\end{align}
where $\bm{\theta}$ is a  $7 \times 1$ vector given by,
\begin{align}
\bm{\theta} & = \begin{bmatrix}
0 & 0 & 0 & 0 & 1 & 1 & 1 \end{bmatrix} ^{\top},
\end{align}
with the mean AoI written as,
\begin{align}
\mathbb{E} [\Delta] & = \frac{\bm{\sigma} {\bm{S}^{-2}} \bm{\theta}}{-\bm{\sigma} \bm{S}^{-1} \bm{\theta}}.  \label{mean_aoi}
\end{align}

For ZW, the matrix structure of $\bm S$ allows one to explicitly write the matrix inverse $\bm{S}^{-1}$ as,
\begin{align} 
\bm{S}^{-1}=-\frac{1}{\mu_1\!+\!\mu_2} & 
\left[
\begin{array}{ccccccc}
1 & \mu_2' & 0 & 0 & 2\mu_1' \mu_2' & \mu_1' & 0 \\
    0 & 1 & 0 & 0 & \mu_1' & 0 & 0 \\
    0 & 0 & 1 & \mu_1' & 2\mu_1'\mu_2' & 0 & \mu_2' \\
    0 & 0 & 0 & 1 & \mu_2' & 0 & 0 \\
    0 & 0 & 0 & 0 & 1 & 0 & 0 \\
    0 & 0 & 0 & 0 & \mu_2' &1 &0 \\
    0 & 0 &0 &0 & \mu_1' & 0 & 1
\end{array}
\right], \label{big2} 
\end{align}
where $\mu_i'=\mu_i/(\mu_1 + \mu_2)$, $i=1,2$, which consequently gives the following closed form expressions for 
$\mathbb{E} [\Phi]$ and $\mathbb{E} [\Delta]$,
\begin{align}
 \mathbb{E} [\Phi] &  =  \frac{2(\mu_1 + \mu_2)}{\mu_1^2 + \mu_1 \mu_2 + \mu_2^2},\label{exp:ZWPAoI} \\
 \mathbb{E} [\Delta] &  = \frac{2(\mu_1^2 + 3\mu_1 \mu_2 + \mu_2^2)}{(\mu_1 + \mu_2)^3}, \label{exp:ZWAoI}
\end{align}
and the latter expression for mean AoI overlaps with the result obtained in \cite{chen_etal_comlet23} for mean AoI for the ZW policy. We note that the distribution, and hence the higher order moments of AoI, and additionally of PAoI, are further obtained for the ZW policy, with the AMC method presented in this paper.

\subsection{Freeze/Preempt (F/P) Policy}
For the first step of the AMC method for the F/P policy, we construct an AMC $Z(t)$ with two absorbing states, similar to the case of the ZW policy, which starts operation at time $t=0$ (corresponding to $t=t_l$ in Fig.~\ref{fig:samplepath}) with the generation of an arbitrary packet, say $P_{\ast}$. 
The transient and absorbing states (resp. transition rates) of the proposed AMC are given in Table~\ref{tab:AMCofFP} (resp. Table~\ref{tab:TRofFP}).
IF (in freeze) states are the ones that a new transmission cannot be initiated, whereas NIF (not in freeze) states are the ones a transmission can be initiated upon availability of either of the two servers. An IF state is visited after the F/P policy forces a freeze upon a new transmission in which case we need to keep track of the phase $\ell$ of the Erlang-$k$ freeze duration which gives rise to the transient states $(j,\ell), \ j \in \mathcal{F}=\{1,2,4,6,8,10,11,12,13 \}$ in which the system is in a freeze period modeled with an Erlang-$k$ distribution with mean $\lambda^{-1}$ and order $k$. This is in contrast to ZW where we did not have any freeze states.

\begin{table}[tb]
\caption{Transient and absorbing states of the AMC process $Z(t)$ for the GAW-2 system employing F/P policy. IF (resp. NIF) stands for in freeze (resp. not in freeze).}
\centering
\begin{tabular}{|c|c|} 
 \hline
 State & Description \\
 \hline \hline
 $(1,\ell), \ 1 \leq \ell \leq k$ & $P_{\ast}$ on $S_1$, $S_2$ idle, IF \\
 \hline
 $(2,\ell), \ 1 \leq \ell \leq k$ & $P_{\ast}$ on $S_2$, $S_1$ idle, IF \\
 \hline
 $3$ & $P_{\ast}$ on $S_1$, $S_2$ busy, $T_1 < T_2$, NIF\\
 \hline
 $(4,\ell), \ 1 \leq \ell \leq k$ & $P_{\ast}$ on $S_1$, $S_2$ busy, $T_1 < T_2$, IF\\
 \hline
 $5$ & $P_{\ast}$ on $S_1$, $S_2$ busy, $T_1 > T_2$,  NIF\\
 \hline
 $(6,\ell), \ 1 \leq \ell \leq k$ & $P_{\ast}$ on $S_1$, $S_2$ busy, $T_1 > T_2$, IF\\
 \hline
 $7$ & $P_{\ast}$ on $S_2$, $S_1$ busy, $T_1 > T_2$, NIF\\
 \hline
 $(8,\ell), \ 1 \leq \ell \leq k$ & $P_{\ast}$ on $S_2$, $S_1$ busy, $T_1 > T_2$, IF\\
 \hline
 $9$ & $P_{\ast}$ on $S_2$, $S_1$ busy, $T_1 < T_2$,  NIF\\
 \hline
 $(10,\ell), \ 1 \leq \ell \leq k$ & $P_{\ast}$ on $S_2$, $S_1$ busy, $T_1 < T_2$, IF\\
 \hline
 $(11,\ell), \ 1 \leq \ell \leq k$ & Both servers are idle, IF\\
 \hline
 $(12,\ell), \ 1 \leq \ell \leq k$ & $P_1$ on $S_1$, $S_2$ idle, IF \\
 \hline
 $(13,\ell), \ 1 \leq \ell \leq k$ & $P_2$ on $S_2$, $S_1$ idle, IF \\
 \hline
 $14$ & $P_1$ on $S_1$, $P_2$ on $S_2$ \\
 \hline
 $15$ & Successful absorbing state \\
 \hline
 $16$ & Unsuccessful absorbing state  \\
 \hline
\end{tabular}
\label{tab:AMCofFP}
\end{table}

\begin{table}[tbh]
\caption{Transition rates for the AMC process $Z(t)$ for the GAW-2 system employing F/P policy.}
\centering
\begin{tabular}{|c|c|c|} 
 \hline
 \multicolumn{3}{|c|}{Transition Rates}  \\
 \hline \hline
 From & To  & Value \\ 
 \hline \hline
 $(1,\ell)$ & $(1,\ell+1)$ when $\ell < k$ & $k \lambda$  \\ \cline{2-3} 
   & $(4,1)$ when $\ell=k$ & $k\lambda$  \\ \cline{2-3} 
   & $(11,\ell)$ & $\mu_1$ \\ \hline 
 $(2,\ell)$ & $(2,\ell+1)$ when $\ell < k$ & $k \lambda$  \\ \cline{2-3} 
   & $(8,1)$ when $\ell=k$ & $k\lambda$  \\ \cline{2-3} 
   & $(11,\ell)$ & $\mu_2$ \\ \hline 
   $3$ & $14$ & $\mu_1$ \\ \cline{2-3} 
       & $16$ & $\mu_2$ \\ \hline
  $(4,\ell)$ & $(4,\ell+1)$ when $\ell < k$ & $k \lambda$  \\ \cline{2-3} 
   & $3$ when $\ell=k$ & $k\lambda$  \\ \cline{2-3} 
   & $(13,\ell)$ & $\mu_1$ \\ \cline{2-3}
   & $16$ & $\mu_2$ \\ \hline
   $5$ & $(12,1)$ & $\mu_1$ \\ \cline{2-3} 
       & $(4,1)$ & $\mu_2$ \\ \hline
   $(6,\ell)$ & $(6,\ell+1)$ when $\ell < k$ & $k \lambda$  \\ \cline{2-3} 
   & $5$ when $\ell=k$ & $k\lambda$  \\ \cline{2-3} 
   & $(11,\ell)$ & $\mu_1$ \\ \cline{2-3}
   & $(1,\ell)$ & $\mu_2$ \\ \hline 
   $7$ & $16$ & $\mu_1$ \\ \cline{2-3} 
       & $14$ & $\mu_2$ \\ \hline
  $(8,\ell)$ & $(8,\ell+1)$ when $\ell < k$ & $k \lambda$  \\ \cline{2-3} 
   & $(7,1)$ when $\ell=k$ & $k\lambda$  \\ \cline{2-3} 
   & $16$ & $\mu_1$ \\ \cline{2-3}
   & $(12,\ell)$ & $\mu_2$ \\ \hline
   $9$ & $(8,1)$ & $\mu_1$ \\ \cline{2-3} 
       & $(12,1)$ & $\mu_2$ \\ \hline
   $(10,\ell)$ & $(10,\ell+1)$ when $\ell < k$ & $k \lambda$  \\ \cline{2-3} 
   & $(9,1)$ when $\ell=k$ & $k\lambda$  \\ \cline{2-3} 
   & $(2,\ell)$ & $\mu_1$ \\ \cline{2-3}
   & $(12,\ell)$ & $\mu_2$ \\ \hline
   $(11,\ell)$ & $(11,\ell+1)$ when $\ell < k$ & $k \lambda$  \\ \cline{2-3} 
   & $(12,1)$ when $\ell=k$ & $k\lambda$  \\ \hline
   $(12,\ell)$ & $(12,\ell+1)$ when $\ell < k$ & $k \lambda$  \\ \cline{2-3} 
   & $14$ when $\ell=k$ & $k\lambda$  \\ \cline{2-3}
   & $15$ & $\mu_1$ \\ \hline
   $(13,\ell)$ & $(13,\ell+1)$ when $\ell < k$ & $k \lambda$  \\ \cline{2-3} 
   & $14$ when $\ell=k$ & $k\lambda$  \\ \cline{2-3}
   & $15$ & $\mu_2$ \\ \hline
   $14$ & $15$ & $\mu_1 + \mu_2$ \\ \hline
\end{tabular}
\label{tab:TRofFP}
\end{table}

The explanation for the transition rates related to freezing are as follows:
\begin{itemize}
\item When $Z(t)$ is in state $(j,\ell), 1 < \ell < k, \ j \in \mathcal{F}$, then a transition to $(j,\ell+1)$ occurs with intensity $k \lambda$ and the phase of the freeze period is incremented by one. 
\item When in state $(j,k),\ j \in \mathcal{F}$, the freeze period ends with intensity $k \lambda$. In this situation, when there are no idle servers, i.e., 
$j \in \{4,6,8,10 \}$, then a state transition from the IF state $(j,k)$ to the corresponding NIF state $j-1$, takes place. 
When there are idle servers, then a fresh packet will be generated and will be placed on the free server (on server 1 if both are idle). While doing so, a new freeze period is initiated and the phase of the freeze period is set to one. These observations lead the CTMC $Z(t)$ to transition from state $(1,k)$ to $(4,1)$, from state 
$(2,k)$ to $(8,1)$, and from state $(11,k)$ to $(12,1)$. 
\item The transitions to state $14$ are from states $(12,k)$ and $(13,k)$, at which there is no need to keep track of the phase of the freeze period since the successful absorbing state is reached upon a service completion from either of the two servers at this state. 
\end{itemize}
We explain the transition rates related to absorption into either one of the two absorbing states $15$ and $16$ as follows:
\begin{itemize}
\item When $Z(t)$ is visiting state $3$ or $(4,\ell)$, $P_{\ast}$ is on $S_1$, $S_2$ is busy transmitting a fresher packet item than $P_{\ast}$ at which a service completion at $S_2$ occurs with intensity $\mu_2$. In this case, $P_{\ast}$ is discarded at the source leading to absorption into state $16$.
\item  When $Z(t)$ is visiting state $7$ or $(8,\ell)$, $P_{\ast}$ is on $S_2$, $S_1$ is busy transmitting a fresher packet item than $P_{\ast}$ at which a service completion at $S_1$ occurs with intensity $\mu_1$. Then, $P_{\ast}$ is discarded and $Z(t)$ is absorbed into state $16$.
\item In states $(12,\ell)$, $(13,\ell)$, and $14$, $P_{\ast}$ was already successfully transmitted, and with transition intensities, $\mu_1$, $\mu_2$ and $\mu_1+\mu_2$, respectively, the packet following $P_{\ast}$ is to be successfully transmitted leading to absorption into state $15$. 
\end{itemize}

We explain the transition rates for the NIF states:
\begin{itemize}
    \item 
    At NIF state $3$ (resp. state $7$), $S_1$ (resp. $S_2$) completes its transmission of $P_{\ast}$ with intensity $\mu_1$ (resp. $\mu_2$) and the timestamp of $P_{\ast}$ is smaller than the other packet in transmission. Thus, $S_2$ (resp. $S_1$) continues transmitting its packet, and a fresh packet is transmitted on the free server $S_1$ (resp. $S_2$) leading to a state transition to state $14$.
    \item 
    At NIF state $5$ (resp. state $9$), $S_1$ (resp. $S_2$) completes its transmission of $P_{\ast}$ with intensity $\mu_1$ (resp. $\mu_2$) and since the timestamp of $P_{\ast}$ is larger than the other packet in transmission, packet on $S_2$ (resp. $S_1$) is discarded, both servers become idle, and a fresh packet is thus transmitted on the free server $S_1$ leading to a state transition to state $(12,1)$ in both cases.
   \item At NIF state $5$, $S_2$ completes the transmission with intensity $\mu_2$ and a new transmission on $S_2$ is immediately initiated which leads a transition to state $(4,1)$. Similarly, at NIF state $9$, $S_1$ completes the transmission with intensity $\mu_1$ and a new transmission on $S_1$ is immediately initiated which leads a transition to state $(8,1)$. 
\end{itemize}
Using Table~\ref{tab:TRofFP}, at this point, we have obtained the generator $\bm Q$ of size $9k+7$
where the north-west block of size $9k+5$ is denoted by $\bm{S}$, the $9k+5 \times 2$ north-east block is denoted by $\bm{V}$, $\bm{V_i}, i=1,2$ denoting the $i$th column of $\bm{V}$, in line with the general form of the generator given in Eqn.~\eqref{eq:Smatrix}.

In the second step of the algorithm, we need to obtain $1 \times 9k+5$ initial probability vector $\bm{\beta}$ of the AMC $Z(t)$. However, one cannot write this vector directly as was the case for ZW. For this purpose, we first construct a recurrent, i.e., no transient states, CTMC $W(t), t \geq 0$ whose state space is very different from that of $Z(t)$. With the construction of $W(t)$ and its steady-state probabilities, we would know exactly the view of the system from the perspective of a newly generated packet.
The states of the recurrent MC (RMC) are given in Table~\ref{tab:StatesofRMC} and its transition rates are given in Table~\ref{tab:TRofRMC}.

\begin{table}[tb]
\caption{states of the RMC $W(t)$ for the GAW-2 system employing F/P policy. IF (resp. NIF) stands for in freeze (resp. not in freeze).}
\centering
\begin{tabular}{|c|c|} 
 \hline
 State & Description \\
 \hline \hline
 $(1,\ell), \ 1 \leq \ell \leq k$ & both servers are idle \\
 \hline
 $(2,\ell), \ 1 \leq \ell \leq k$ & $S_1$ busy, $S_2$ idle \\
 \hline
 $(3,\ell), \ 1 \leq \ell \leq k$ & $S_1$ idle, $S_2$ busy \\
 \hline
$(4,\ell), \ 1 \leq \ell \leq k$ & $P_1$ on $S_1$, $P_2$ on $S_2$, $T_1 < T_2$, IF\\
 \hline
 $(5,\ell), \ 1 \leq \ell \leq k$ &  $P_1$ on $S_1$, $P_2$ on $S_2$, $T_1 > T_2$, IF\\
 \hline
$6$ & $P_1$ on $S_1$, $P_2$ on $S_2$, $T_1 < T_2$, NIF\\
 \hline
 $7$ & $P_1$ on $S_1$, $P_2$ on $S_2$, $T_1 > T_2$, NIF \\
 \hline
\end{tabular}
\label{tab:StatesofRMC}
\end{table}

\begin{table}[tb]
\caption{Transition rates for the RMC process $W(t)$ for the GAW-2 system employing F/P policy.}
\centering
\begin{tabular}{|c|c|c|} 
 \hline
 \multicolumn{3}{|c|}{Transition Rates}  \\
 \hline \hline
 From & To  & Value \\ 
 \hline \hline
 $(1,\ell)$ & $(1,\ell+1)$ when $\ell < k$ & $k \lambda$  \\ \cline{2-3} 
   & $(2,1)$ when $\ell=k$ & $k\lambda$  \\ \hline 
$(2,\ell)$ & $(2,\ell+1)$ when $\ell < k$ & $k \lambda$  \\ \cline{2-3} 
   & $(4,1)$ when $\ell=k$ & $k\lambda$  \\ \cline{2-3} 
   & $(1,\ell)$ & $\mu_1$ \\ \hline 
  $(3,\ell)$ & $(3,\ell+1)$ when $\ell < k$ & $k \lambda$  \\ \cline{2-3} 
   & $(5,1)$ when $\ell=k$ & $k\lambda$  \\ \cline{2-3}
   & $(1,\ell)$ & $\mu_2$ \\ \hline
   $(4,\ell)$ & $(4,\ell+1)$ when $\ell < k$ & $k \lambda$  \\ \cline{2-3} 
   & $6$ when $\ell=k$ & $k\lambda$  \\ \cline{2-3} 
   & $(3,\ell)$ & $\mu_1$ \\ \cline{2-3}
   & $(1,\ell)$ & $\mu_2$ \\ \hline 
  $(5,\ell)$ & $(5,\ell+1)$ when $\ell < k$ & $k \lambda$  \\ \cline{2-3} 
   & $7$ when $\ell=k$ & $k\lambda$  \\ \cline{2-3} 
   & $(1,\ell)$ & $\mu_1$ \\ \cline{2-3}
   & $(2,\ell)$ & $\mu_2$ \\ \hline
   $6$ & $(5,1)$ & $\mu_1$ \\ \cline{2-3} 
       & $(2,1)$ & $\mu_2$ \\ \hline
   $7$ & $(2,1)$ & $\mu_1$ \\ \cline{2-3} 
       & $(4,1)$ & $\mu_2$ \\ \hline
\end{tabular}
\label{tab:TRofRMC}
\end{table}

Now, let $\pi(j,\ell), 1 \leq \ell \leq k$ (resp. $\pi(i),i=6,7$) denote the steady-state probability of being in state $(j,\ell)$ (resp. state $i$), at an arbitrary time. These probabilities can be found from the stationary probabilities of the generator $\bm{P}$ with size $5k+2$ constructed out of the transition rates given in Table~\ref{tab:TRofRMC}. Let $f$ denote the overall intensity of new packets joining the GAW-2 system. It is not difficult to see that new packet arrivals can only be generated at NIF states, and therefore,
\begin{align}
    f &= k \lambda \sum_{j=1}^3 \pi(j,k)  + (\mu_1 + \mu_2) \sum_{i=6}^7 \pi(i).
\end{align}

Now, we are ready to link the RMC $W(t)$ to the AMC $Z(t)$. For this purpose, we need to find out which state of the AMC $Y(t)$ a new packet $P_{\ast}$ will be at, just after joining the GAW-2 system. We observe the following:
\begin{itemize}
    \item Let the probability $p_1$ be defined as $p_1=(k\lambda \pi(1,k) + \mu_2 \pi(6) + \mu_1 \pi(7))/f$. With probability $p_1$, the new packet $P_{\ast}$ will join the system at state $(1,1)$ of the AMC $Z(t)$.
   \item  Let the probability $p_2$ be defined as $p_2=(k\lambda \pi(2,k) + \mu_2 \pi(7))/f$. With probability $p_2$, the new packet $P_{\ast}$ will join the system at state $(10,1)$ of the AMC $Z(t)$.
   \item  Let the probability $p_3$ be defined as $p_3=(k\lambda \pi(3,k) + \mu_1 \pi(6))/f$. With probability $p_3$, the new packet $P_{\ast}$ will join the system at state $(6,1)$ of the AMC $Z(t)$.
\end{itemize}
Based on these observations, the initial probability vector $\bm{\beta}$ is found as a row vector of zeros except for the entries corresponding to states $(1,1)$, $(10,1)$, and $(6,1)$ (of the AMC $Z(t)$)
with the values $p_1$, $p_2$, and $p_3$, respectively.

\begin{figure*}[t]
    \centering
    \includegraphics[width=0.85\textwidth]{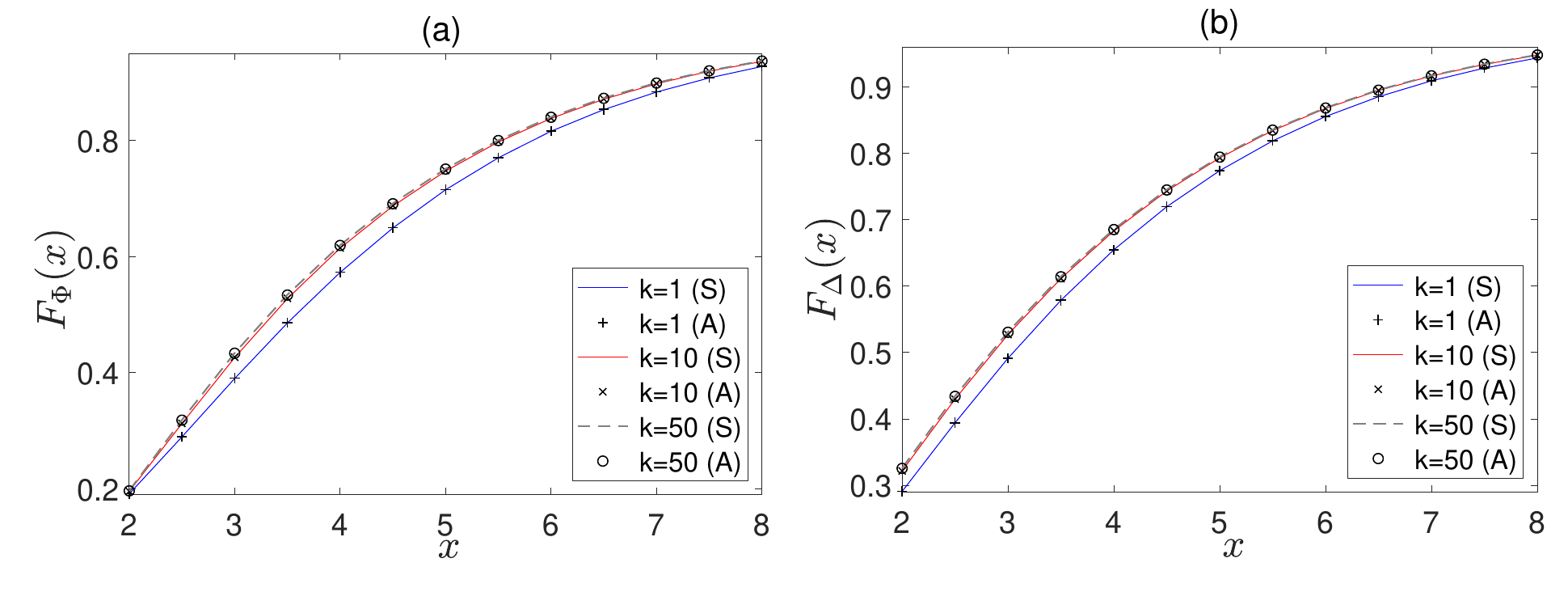}
    \caption{(a) CDF of PAoI, $F_{\Phi}(x)$, (b) CDF of AoI, $F_{\Delta}(x)$, depicted as a function of $x$ for Erlang-$k$ distributed freeze time with $k \in \{ 1,10,50 \}$. (S) and (A) refer to results obtained with simulations and the analytical method, respectively.}
    \label{fig:example1}
\end{figure*}

At the beginning of the final step, we have obtained the characterizing matrices $\bm{\beta}$, $\bm{S}$, and $\bm{V}_1$ of the AMC $Z(t)$. 
We observe that the dashed black curve in Fig.~\ref{fig:samplepath} starting at $t_l$ until $d_{l+1}$ amounts to the time spent before absorption for the AMC $Z(t)$ in successful absorption cycles. The portion of this curve from $t_l$ until $d_{l}$ is spent in states $(1,\cdot)$-$(10,\cdot)$ at which $P_{\ast}$ is always present and this portion of the curve does not overlap with the AoI curve. On the other hand, the portion of this curve from $d_l$ until $d_{l+1}$ spent in states  $(11,\cdot)$-$(13,\cdot)$ and state $14$ is the same as the AoI curve in cycle-$l$. Similar to the analysis of ZW, we construct a column vector $\bm{h}$ being all zeros except for a unit entry corresponding to the states $(j,\ell), 11 \leq j \leq 13, 1 \leq \ell \leq k$, and also the state $14$. Finally, the density of peak AoI, the mean peak AoI, the density of AoI, and the mean AoI, are given by the same expressions \eqref{density_paoi},\eqref{mean_paoi},\eqref{density_aoi} and \eqref{mean_aoi}, respectively, which were obtained for the ZW policy. We omit the proofs which are identical to those presented for ZW once the AMC $Z(t)$ is constructed.

\section{Numerical Examples} \label{sec:Numerical}
In the first numerical example, we validate the analytical model developed for the F/P policy by simulations. For this purpose, we fix $\mu_1=0.5, \mu_2=0.1, \lambda=1$. For three values of the parameter $k \in \{ 1,10,50 \}$, we employ Erlang-$k$ freezing and using the analytical model (A) and simulations (S), we obtain the cdf of PAoI and AoI, namely $F_{\Phi}(x)$ and $F_{\Delta}(x)$, respectively, which are depicted in Fig.~\ref{fig:example1} which shows that the analytical and simulation results are in perfect agreement. 
Moreover, the PAoI and AoI distributions obtained with $k=10$ and $k=50$ are very close which shows that an Erlang-10 freezing time can quite accurately be used to obtain an approximation of the actual AoI distribution with deterministic freezing in dual status update systems. 

\begin{figure}[t]
    \centering
    \includegraphics[width=0.48\textwidth]{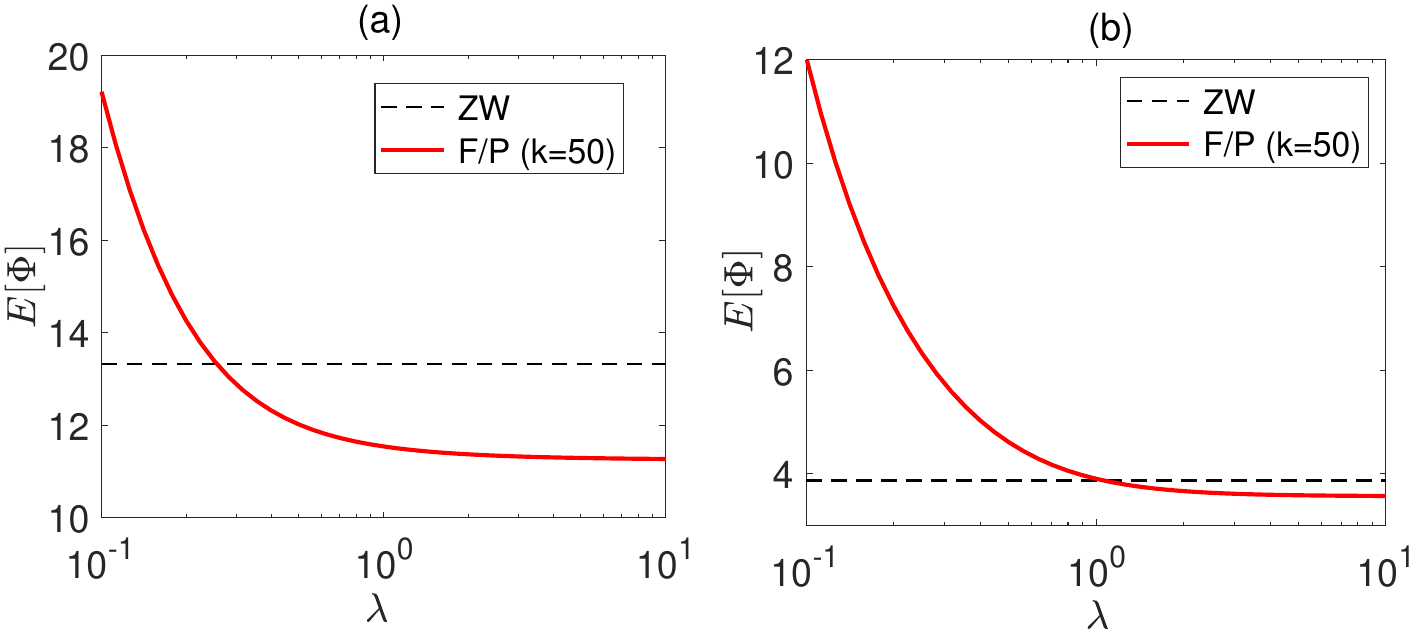}
      \vspace*{-0.3cm}
    \caption{Mean PAoI obtained with F/P policy as a function of $\lambda$ when $\mu_2=0.1$ and (a) $\mu_1=0.1$ (b) $\mu_1=0.5$. The mean PAoI for the ZW policy obtained with \eqref{exp:ZWPAoI} is also depicted as reference.}
    \label{fig:example2}
\end{figure}

\begin{figure}[t]
    \centering
    \includegraphics[width=0.48\textwidth]{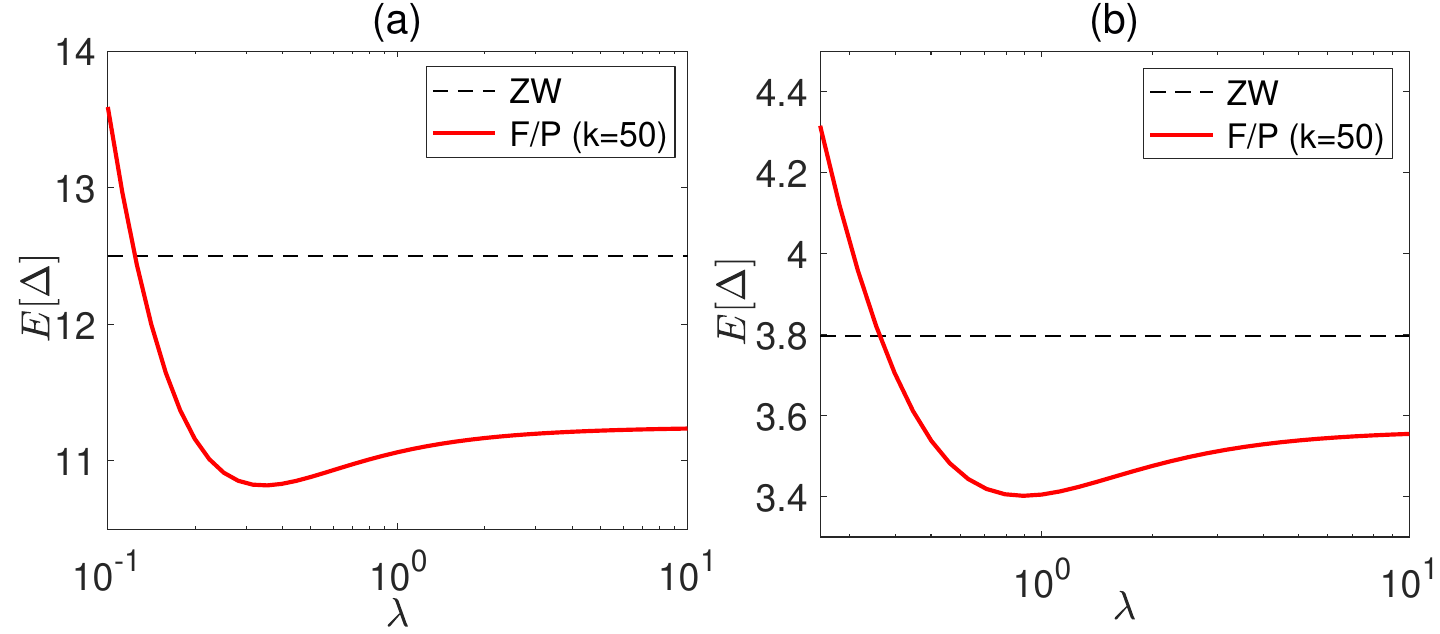}
      \vspace*{-0.3cm}
    \caption{Mean AoI obtained with F/P policy as a function of $\lambda$ when $\mu_2=0.1$ and (a) $\mu_1=0.1$ (b) $\mu_1=0.5$. The mean AoI for the ZW policy obtained with \eqref{exp:ZWAoI} is also depicted as reference.}
    \label{fig:example3}
\end{figure}

In the second numerical example, we fix $k=50$ and study the performance of the F/P policy in terms of mean PAoI and mean AoI as a function of the freezing rate $\lambda$. As in the previous example, we fix $\mu_2=0.1$. The mean PAoI (resp. mean AoI) is plotted in Fig.~\ref{fig:example2} (resp. Fig.~\ref{fig:example3}) for the F/P policy as a function of $\lambda$ for two values of $\mu_1 \in \{ 0.1, 0.5 \}$. The results show that freezing does not improve the mean PAoI performance and needs to be avoided if one is interested in the minimization of mean PAoI only. However, preemption-only policy (obtained from F/P as $\lambda \rightarrow \infty$) is shown to be beneficial in terms of mean PAoI in comparison to ZW, especially when the service rates of the two servers are close to each other. This situation is different when the mean AoI is taken as the performance metric; see Fig.~\ref{fig:example3}.  For small values of $\lambda$, the mean freezing time is long and therefore the servers are not efficiently utilized leading to a rise in mean AoI. However, there appears to be a value of $\lambda$, called $\lambda^{\ast}$, at which the mean AoI takes its minimum value. When $\lambda > \lambda^{\ast}$, the mean AoI rises to its asymptotic value as $\lambda \rightarrow \infty$, i.e., freezing is abandoned. Therefore, with proper choice of the mean freezing time, F/P outperforms ZW not only with preemption at the source feature, but also with freezing.
Moreover, mean AoI appears to be a unimodal function (see \cite{AppliedNumericalAnalysis} of the freezing rate $\lambda$ which means that there is exactly one point $\lambda^*$ on the interval $\lambda \in (0,\infty)$ for which the mean AoI takes its minimum value and when $\lambda < \lambda^*$ (resp. $\lambda > \lambda^*$), mean AoI is a strictly decreasing function (resp. strictly increasing function) of $\lambda$. There are efficient search algorithms for finding the minimum of unimodal functions such as the Golden section search algorithm \cite{NumericalRecipes,AppliedNumericalAnalysis} which does not require the use of derivatives. 

\begin{figure}[t]
    \centering
    \includegraphics[width=0.35\textwidth]{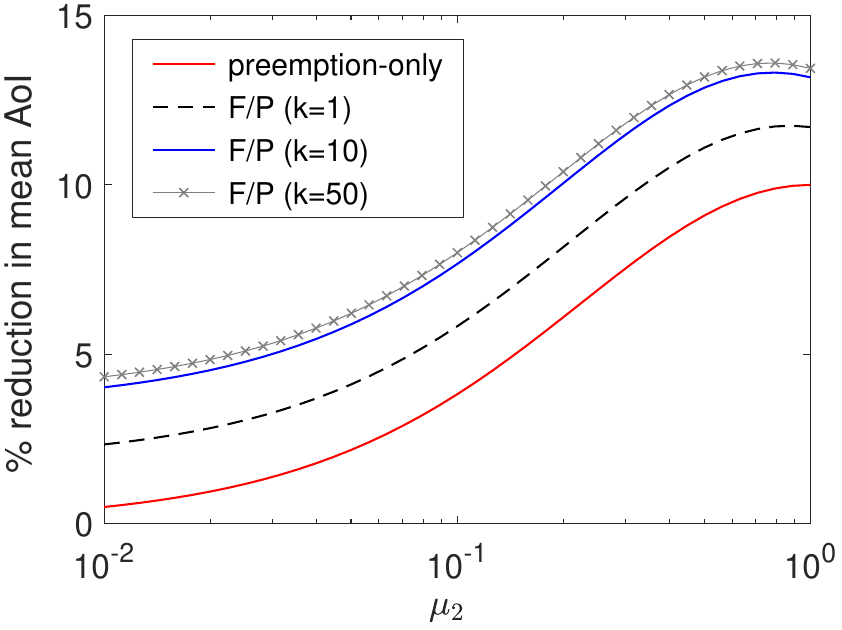}
      \vspace*{-0.3cm}
    \caption{Percentage reduction in mean AoI as a function of $\mu_2$, obtained by the preemption-only policy and F/P policy for Erlang-$k$ distributed freeze time with $k \in \{ 1,10,50 \}$, with respect to the baseline ZW policy.}
    \label{fig:example4}
\end{figure}

\begin{figure}[t]
    \centering
    \includegraphics[width=0.37\textwidth]{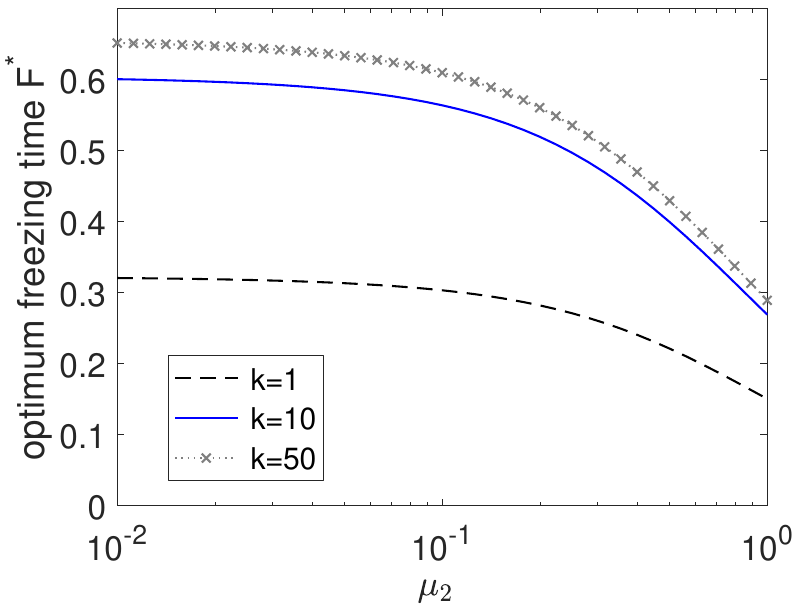}
    \vspace*{-0.3cm}
    \caption{Optimum freezing time $F^*$ depicted as a function of $\mu_2$ for three values of $k \in \{ 1,10,50 \}$ }
    \label{fig:example5}
\end{figure}

In the final numerical example, for the case of $\mu_1=1$, we compare ZW against the preemption-only policy, i.e., freezing disabled, and also with Erlang-$k$ freezing for three choices of $k \in \{ 1,10,50 \}$ each employing the optimum freezing rate $\lambda^{\ast}$ using the analytical model developed in this paper along with the Golden section search algorithm detailed in \cite{NumericalRecipes,AppliedNumericalAnalysis}. 
The optimum mean freezing time is denoted by $F^*$ which is the reciprocal of $\lambda^{\ast}$. 
Mean AoI is obtained for the preemption-only policy by setting $k=1$ and a very large freezing rate, i.e., $\lambda = 10^8$, without having to construct a separate AMC-based model for this specific scenario.
For each of the studied policies, we compute the percentage reduction in mean AoI with respect to the ZW policy for varying values of $\mu_2 \in [0.01, 1]$. The results are depicted in Fig.~\ref{fig:example4}. We have the following observations. We observe that a reduction in mean AoI of up to $\%$ 10 is possible with the preemption-only policy. By enabling freezing using the mean freezing time $F^*$, larger reductions in mean AoI up to $\%$ 13.60 (attained when $\mu_2$ equals 0.7943) are possible for larger choices of $k$, which leads us to conclude that deterministic freezing appears to be more advantageous than random freezing.  The optimum freezing time $F^*$ is plotted in Fig.~\ref{fig:example5} as a function of $\mu_2$ for the three F/P policies corresponding to $k \in \{1,10,50 \}$. We observe that $F^*$ increases when heterogeneity increases, i.e., $\mu_2 \ll \mu_1$. On the other hand, when $\mu_1=\mu_2$, $F^*=0.2894$ for $k=50$ which is significantly different than zero.

\section{Conclusions and Future Work} \label{sec:Conclusions}
We propose a non-work conserving F/P (freeze/preempt) policy for a status update system with one information source and dual servers. 
For the conventional ZW (zero wait) policy as well as F/P policy with Erlang-$k$ distributed freezing under the assumption of exponentially distributed service times, we propose an analytical model based on absorbing Markov chains to exactly obtain the distributions of AoI and PAoI. 
We validate the analytical model with simulations. We also show that freezing is not advantageous for mean PAoI. However, freezing can be quite beneficial for mean AoI reduction with the appropriate choice of the freezing time, which is shown to be possible thanks to the developed analytical model in conjunction with the use of the Golden section search algorithm. It is also shown that preemption at the source is beneficial for both mean PAoI and AoI, in contrast to discarding at the monitor, a feature that belongs to the ZW policy. Future work could investigate the possibility of developing analytical models using absorbing Markov chains, for status update systems involving more than two servers, packet errors, non-exponentially distributed service times, and random arrivals. Another future extension may involve the study of multiple sources and multiple servers.  


\end{document}